\documentclass[manuscript]{aastex}

\slugcomment{Not to appear in Nonlearned J., 45.}

\shorttitle{VLBI Monitoring of 3C 66B at 22 GHz}
\shortauthors{Sudou, Iguchi, and Zhao}

\begin{document}


\title{VLBI Monitoring of the Sub-parsec-scale Jet in the Radio Galaxy 3C 66B at 22 GHz}


\author{Hiroshi Sudou}
\affil{Faculty of Engineering, Gifu University, 1-1 Yanagido, Gifu 501-1193, Japan; sudou@gifu-u.ac.jp}

\author{Satoru Iguchi}
\affil{National Astronomical Observatory of Japan, 2-2-1 Osawa, Mitaka, Tokyo 181-8588, Japan; s.iguchi@nao.ac.jp}

\and

\author{Guang-Yao Zhao}
\affil{Korea Astronomy and Space Science Institute, Daejeon 305-348, Korea; gyzhao@kasi.re.kr}






\begin{abstract}

We present measurements of the proper motion of the sub-parsec scale jet at 22 GHz in the nearby FR I galaxy 3C 66B.
Observations were made using the VLBA at six epochs over four years. 
A phase-referencing technique was used to improve the image quality of the weak and diffuse jet components. 
We find that the inner knots are almost stationary, though one of them was expected to be detected with an apparent speed of 0.2 mas yr$^{-1}$ according to 8 GHz monitoring at the same observation epochs.
{  Clear flux variations are not observed in the core at 22 GHz; in contrast, clear flux enhancement is observed in the core at 8 GHz.
We discuss a possible explanation: If the jet has helical structure, the viewing angle of the jet at 8 and 22 GHz differ by a few degrees 
if the jet direction is almost along our line of sight. 
Although these results may imply the existence of a two-zone jet, which has been suggested in certain radio galaxies, it cannot explain 
the fact that the jet at the higher frequency is slower than that at the lower frequency. 
}

\end{abstract}


\keywords{ galaxies : individual (3C 66B) -- galaxies : active -- galaxies : jets -- galaxies : nuclei
}



\section{Introduction}

Many efforts to clarify the radio properties of active galactic nuclei (AGNs) have advanced our understanding of relativistic jets \citep{zensus97, ghisellini93}.
Recent MOJAVE activities are helpfull in understanding of the statistical properties of radio jets \citep{lister13, homan15}.
They examined the complete AGN data-set using the VLBA 2 cm survey, which contains many compact sources such as BL Lac objects, quasars, and bright radio galaxies \citep{lister05}. 
This samlple reveals the detailed kinematics of highly relativistic jets based on the observed superluminal speeds and accelerating motion. 
{ 
Owing to their low radio luminosities, only a handful of FR I radio galaxies have been observed using very long baseline interferometry (VLBI), 
and even fewer have been observed with VLBI at multiple epochs.
Here we present the results of multi-epoch VLBI observations of the FR I radio galaxy 3C 66B. 

In 3C 66B ($z\sim0.02$), which is hosted by the giant radio galaxy UGC 01841, we can find a strong radio jet and counterjet out to 100 kpc.
The jets have been well observed in the centimeter-wave \citep{hardcastle96}, millimeter-wave \citep{iguchi10}, infrared \citep{tansley00}, optical \citep{macchetto91}, and X-ray \citep{hardcastle01} regimes. 
High-resolution imaging of 3C 66B using VLBI has revealed the properties of the inner jet such as its structure and polarization. 
The main findings are as follows: 
 (1) The possible presence of a faint counterjet in 3C 66B was revealed at both 5 GHz \citep{giovannini01} and 8 GHz \citep{kharb09}. 
In an analysis of the 5 GHz image, the intensity ratio between the jet and counterjet ($R$) was estimated to be approximately 10 within a distance of $\sim 1$ pc from the core, 
and $R$ became larger than 100 downstream along the jet \citep{giovannini01}. 
This fact indicates possible acceleration of the jet outflow on a parsec scale.
(2) Polarization was detected along the jet edge within $\sim$5 pc. 
The electric vectors in the inner jet on scales of 0.3 pc are parallel to the jet direction, which implies a transverse magnetic field like that typical of BL Lac objects \citep{kharb05}. 
On the other hand, on a scale of 5 pc, the electric vectors are transverse to the jet direction, implying an aligned magnetic field \citep{kharb12}. 
(3) Multi-epoch monitoring of the 3C 66B inner jet at 8 GHz  showed that the measured apparent speed ranges from 0.30 $c$ to 0.96 $c$ within a distance of $\sim 1$ pc from the core \citep{sudou11}.
The increase in the apparent jet speed can be interpreted as being caused by either the acceleration or a change in the viewing angle of the jet. 
The presence of accelerating jet components in many other sources has been suggested; for example, approximately one third of the blazars in the MOJAVE sample show a systematic change in the jet speed \citep{homan09, homan15}. 
}

On the basis of these indications, we would like to apply imaging monitoring at higher frequencies because it has the advantage of revealing the region closer to the central engine, 
according to plasma theory of the jet \citep{konigl81, lobanov95}.  
Here, we report on multi-epoch 22 GHz observations of the sub-parsec scale jet in 3C 66B over four years. 
The observations and data reduction are  described in \S 2, and measurements of the position shift of the jet and flux variation in the core are discussed in \S 3.

In this paper, we use a Hubble constant ($H_0$) of 71 km s$^{-1}$ Mpc$^{-1}$ , a matter density ($\Omega_{\rm M}$) of 0.27, and a vacuum energy ($\Omega_{\Lambda}$) of 0.73, 
consequently, an angular size or separation of 1 mas corresponds to 0.436 pc at the distance of 3C 66B.

\section{Observations and Data Reduction}

3C 66B was observed at 22.2 GHz using the VLBA \citep{napier94} of  the  NRAO\footnote{The National Radio Astronomy Observatory is a facility of the National Science Foundation operated under cooperative agreement by Associated Universities, Inc.}  from 2001.20 to 2005.54, at seven epochs. 
Because of bad weather conditions and antenna problems, the absolute flux density data for the second epoch (2001.48) showed very low quality, 
therefore, we excluded these data from further imaging analysis. 
Consequently, we obtained six sets of data for each frequency.  
Details of the observations for each epoch are listed in Table 1. 
In addition, 8 GHz observations were also conducted at the same epochs as the 22 GHz observations \citep{sudou11}. 

Phase-referencing observations were performed to improve both the image quality and the astrometric accuracy. 
In this paper, we focus on the improvement in the image quality, and we will present the astrometric results in a separate paper. 
The 22 GHz observations consisted of nodding-mode observations of 3C 66A and 3C 66B with a switching cycle of 30 s. 
3C 66A is a very bright quasar exhibiting a compact core and one-sided southward jet \citep{taylor96, jorstad01, cai07, zhao14}. 
Since this source is very close in the sky to 3C 66B ($\sim 0.1$ degree, it is a very good source for phase and amplitude calibration of 3C 66B.

The data were recorded in the VLBA format with four intermediate frequency bands with a bandwidth of 8 MHz each; thus the total bandwidth was 32 MHz. 
The total bit rate for both frequencies was 128 Mbps using a 2-bit sampling mode.
The observed data were correlated with the VLBA correlator in Socorro.

The correlated data for 3C 66B and 3C 66A were calibrated using the AIPS package developed by the NRAO.
The data for all epochs were treated in a similar fashion;  the residual phase delays and delay rates, which were due mainly to antenna position errors and atmospheric phase fluctuations were calibrated using global fringe-fitting \citep{schwab83} of the data from 3C 66A with a solution interval of 2 min.
Because 3C 66A and 3C 66B were observed alternately every 15 s with a switching time of 30 s, the effective scan time of 3C 66A is just half of the input solution interval. 
The solutions of the global fringe-fitting were applied to the data of both 3C 66A and 3C 66B. 
The visibility amplitudes were calibrated using the measured system temperature and antenna gain curves. 
Bandpass calibration was performed using 0133+476 for all epochs. 

The images of 3C 66A were made using CLEAN with uniform weighting and a self-calibration technique.  
They are in good agreement with previous results from the same data sets \citep{cai07, zhao14}. 
The final solutions for the amplitude and phase obtained by self-calibration of 3C 66A for a time interval of 0.1 min were applied to the 3C 66B data
to calibrate the results for short period variations due to atmospheric effects.

The calibrated visibility data of 3C 66B were coherently averaged over 30 s using DIFMAP \citep{shepherd97}. 
The images were made with DIFMAP using CLEAN and a self-calibration technique. 
The above process using phase-referencing data greatly improve the image quality for 3C 66B, making it possible to image very weak extended jets, 
as reported by Asaki et al. (2014) for water maser mapping.   

Figure 1 shows the final six-epoch images of 3C 66B at 22 GHz. 
The obtained CLEAN beams differ from each other (see also Table 1) because of data loss at some antennas due to bad weather conditions or antenna problems at each epoch. 
Note that at the epochs 2004.80 and 2005.35 the beam is approximately two times larger than at other epochs, mainly because all data from the St. Croix station were lost. 

\section{Results}

\subsection{Structure of the jet}

The images at 22 GHz show the sub-parsec scale jet structure within a distance of 1.3 pc (3 mas) from the nucleus (Figure 1).
The averaged position angle (PA) of the jet is estimated to be 44 $\pm$ 2 degrees, which differs from that at 8 GHz by $\sim$10 degrees \citep{sudou11}.
A clear bend appears 0.4 pc (0.8 mas) from the core, in particular at epochs 2005.05 and 2005.54. 
This bend was also seen in the 8 GHz images at about 1 pc from the core. 

{ 
We  cannot find a significant counterjet component at any epoch, in contrast with previous lower frequency results \citep{giovannini01, kharb09}. 
The lower limit of the jet-to-counterjet intensity ratio from our 22 GHz image is 3, which is consistent with the reported values. 
}

Figure 2 shows an intensity profile of the jet along the averaged PA. Although the profile seems to be very smooth, a weak component can be found at approximately 0.3 pc (0.6 mas) from the core. 
The time evolution of the jet profile does not indicate clear  systematic changes due to jet motion. 

\subsection{Changes in knot positions}

We applied a single Gaussian fitting to the core and a point source fitting to the jet, which is similar to the method used in the analysis of the 8 GHz results \citep{sudou11}. 
When a circular Gaussian fitting was applied to the jet, the positions of each knot were almost the same as those from the point source fitting.  
We successfully fitted one or two jet components in the 22 GHz images: knots E4 and E5 at 0.3 pc (0.6 mas) and 0.1 pc (0.3 mas) from the core position, $F_{\rm K}$, respectively. 
The labels for the knots are based on the 8 GHz results in terms of their distance from the core. 
Namely, we deduced that knot E4 is likely to be the same as that detected at 8 GHz. 
Knot E5 is closer to the core than knot E4 and was not found in the 8 GHz observations probably because the spatial resolution at 8 GHz was insufficient to resolve this component from the core.  

The model fitting errors were generally estimated from the value that increases $\chi^2$ by $\Delta \chi^2$ from the minimum value. 
For instance, the values of $\Delta \chi^2$ = 1.0 and 6.6 correspond to confidence levels of 1$\sigma$ and 3$\sigma$, respectively,  for a parameter of interest \citep{avni76}. 
The above simple fitting error analysis of the flux density leads to very small errors (typically 1 mJy). 
We added flux errors of 20 \% to the measured integrated flux density according to the reported accuracy of the absolute amplitude  calibration of the VLBA at higher frequencies \citep{cai07}. 
The position errors, with a 3$\sigma$ confidence level, were estimated to be approximately 10 \% of the beam size. 
To be conservative, we applied position errors of 50 \% of the beam size according to typical fitting results for a very weak and diffuse jet observed by the VLBA \citep{bottcher05, jorstad05}.
The fitted models  are given in Table 2. The fitted positions of the knots are denoted by crosses (+) and that of the core is denoted by an ellipse in the images in Figure 1.  

Figure 3 shows the position change of the knots with respect to the core position at each epoch. 
Knot E4 is expected to have a speed of $\sim 0.2$ mas yr$^{-1}$, according to a suggestion from the analysis of the proper motion at 8.4 GHz.
We performed a weighted least-squares fitting of the jet components. 
Assuming that the knots move along a straight line at an equal speed, we derived the fits for the right ascension (R.A.) direction ($X$) and the declination (Decl.) direction ($Y$) separately.
We estimated the proper motion of knot E4 to be 0.02 $\pm$ 0.02 and $-$0.02 $\pm$ 0.01 mas yr$^{-1}$ in the $X$ and $Y$ directions, respectively. 
This measurement means that, contrary to expectation, knot E4 is almost stationary. 
Similarly, the proper motion of knot E5 was estimated to be 0.01$\pm$ 0.01 and 0.03 $\pm$ 0.01 mas yr$^{-1}$ in the $X$ and $Y$ directions, respectively. 
Considering this analysis, we cannot find any significant proper motion of either knot (E4 or E5).

\subsection {Flux variations of the core}

In Figure 4, we compare the flux variations of the core between the 8 and 22 GHz observations. 
Before the core flux was measured, all images at both frequencies were restored with the 1-mas beam to avoid the resolution effect on the flux measurements. 
A clear correlation  cannot be found between these frequencies. 
In two of the six epochs (Epochs 2002.45 and 2005.54), the core flux was enhanced only at 8 GHz, appearing 20 \% larger than the averaged value. 
Such flux enhancement of the compact core is likely to be a signature of emerging jet components (Nagai et al. 2013).

\subsection {Spectral index image between 8 and 22 GHz}

{ 
Figure 5 shows the images at 8 and 22 GHz with the same restoring beam of 0.7 $\times$ 0.7 mas and the spectral index image made from these images 
assuming that the position of the optically thin jet is the same at these two frequencies.
Note that a detailed analysis of the core shift among the observing frequencies showed that the core shift between 8 and 22 GHz is small 
(about 0.1 $-$ 0.2 mas) compared with the beam size \citep{sudou02, zhao15}. 
Both the jet and the core show a steep spectral index ($F\propto\nu^{\alpha}$). 
In the jet region, the highest value is $\alpha\sim-0.4$ at a distance of 0.4 mas from the core,
which almost corresponds to the position of knot E5. 
Then  it becomes gradually steeper along the jet to $\alpha\sim-0.9$ in the downstream direction. 
A clear gradient of the spectral index along the direction transverse to the jet cannot be seen.  
In the core region, $\alpha\sim-0.6$,
since the spectral index at the core between 5 and 8 GHz is inverted \citep{kharb12}, the turn over frequency is indicated to be between 8 and 22 GHz. 
}

\section{Discussion}

Parsec-scale jets in radio galaxies generally tend to show mildly relativistic speeds \citep{ghisellini93,asada14}.
These slow jets can be interpreted as slow pattern speed (SPS) features that could be due to either stationary shocks in the jet or the jet bending effect.
Lister et al. (2013) showed that approximately 25 \% of radio galaxies contain SPS components. 
In 3C 66B, knots E4 and E5 appear as an SPS component at 22 GHz because they are almost stationary. 

{ 
Interestingly, knot E4 showed much faster proper motion at 8 GHz than at 22 GHz at the same observation epochs. 
The first question we consider here is the effect of the motion of the core.
For 3C 66B, the position change of the core has been estimated to be at most 0.04 mas at 8 GHz \citep{sudou03},
and actually the effect of the core motion on measurement of the jet motion has been shown to be limited \citep{sudou11}.
Further, the effect is expected to be much less at 22 GHz than that at 8 GHz (Sudou et al. in preparation). 
Thus, it is reasonable to suppose that the core motions has an almost negligible effect on the analysis of the jet motion.

Multi-frequency monitoring has been applied to a few bright radio galaxies such as the FR I galaxy 
3C 120 \citep{gomez01} and the FR II galaxy Cyg A \citep{boccardi15}.
These observations showed that faster proper motion appeared at the higher frequency than at the lower frequency.
For instance, in 3C 120, the strong component showed superluminal motion at 43 GHz, but the data were compatible with it being stationary at 22 GHz,
although the difference in the spatial resolution between these frequencies was pointed out.
}
This frequency dependence implies that the velocity structure varies depending on the observing frequency.
For example, a two-zone jet model with a fast spine surrounded by a slow layer structure has been proposed 
to explain the co-existence of both strongly relativistic TeV emission and mildly relativistic jets in FR I radio galaxies and BL Lac objects
 (e.g., Chiaberge et al. 2000).
{  
This model was also introduced to explain both the limb brightened structure and very slow motion of a parsec-scale jet, e.g., M 87 \citep{reid89, kovalev07}.
It is natural that observations of the jet at the higher frequency should trace the faster layer (spine) and those at the lower frequency should trace the slower layer (sheath).
However, our proper motion measurement of 3C 66B is not in agreement with this two-zone model; i.e., the jet is faster at the lower frequency. 

We would like to point out that a geometrical effect can explain the difference in the jet speed. 
The proper motion can be very small if the jet is aligned almost along our line of sight (e.g., Urry \& Padovani 1995). 
For example, this effect might explain the result of no apparent motion in the quasar Mrk 231 over 10 yr \citep{reynolds09}. 
Assuming a helical jet structure, we can explain the  difference between the viewing angles at 22 and  8 GHz 
by the difference in the positions of the knots at these frequencies.
The viewing angle of the inner jet at 8 GHz is estimated to be $\sim$ 5 degree, where $\beta\sim0.77$ is estimated from the Doppler factor indicated by the time variation of the core flux at the millimeter wavelengths \citep{iguchi10}.
For example, if the viewing angle of the inner jet at 22 GHz is regarded as 2 degree with the same $\beta$, the proper motion at 22 GHz is less than half of that at 8 GHz
(see Figure 6). 
Finally, we would like to point out that both Mrk 231 and 3C 66B are candidates for supermassive binary black holes (SMBs) \citep{yan15,sudou03} .
In future observational studies of slow jet motion,
it would be very interesting to search for difference between the jet motions of SMBs and non-SMBs.

}

\section{Conclusion}

We presented VLBI monitoring of 3C 66B at 22 GHz at six epochs over four years. 
The proper motion was estimated to be less than 0.03 mas yr$^{-1}$ (0.013 pc yr$^{-1}$, 0.04 $c$) within a distance of 1 pc from the core, suggesting the presence of SPS components. 
The flux at 22 GHz is almost stable within the errors, in contrast with that at 8 GHz. 
These facts indicate the possibility that we observed different parts of the jet depending on the observing frequency. 
{  If the inner jet exhibits helical structure, the viewing angle of the inner jet is different in each region we observed.
The variation in the jet speed depending on the observing frequency could be interpreted as indicating that the inner jet is aligned almost along the line of sight. }

\acknowledgments

We wish to thank an anonymous referee for useful comments.


Facilities: \facility{VLBA}



\begin{deluxetable}{ccccc}
\tablecolumns{5}
\tablewidth{0pc}
\tablecaption{Observations and image parameters.}
\tablehead{
\colhead{Epoch}     &
\colhead{Int. Time} &
\colhead{Beam Size} & 
\colhead{Beam PA}   & 
\colhead{rms noise}\\

\colhead{[yr]}  & 
\colhead{[hr]} &
\colhead{[mas $\times$ mas]} & 
\colhead{[deg]}   & 
\colhead{[mJy/beam]}\\

\colhead{(1)} & 
\colhead{(2)} & 
\colhead{(3)} & 
\colhead{(4)} &
\colhead{(5)} 
}
\startdata

2001.20 	&		1.3 	&	0.40 $\times$ 0.27	&	-8.9  	&	0.7 	\\
2002.45 	&		3.0 	&	0.52 $\times$ 0.28	&	   2.0 	&	0.9 	\\
2004.80 	&		1.0 	&	0.62 $\times$ 0.43	&	  1.2 	&	0.7 	\\
2005.05 	&		1.5 	&	0.40 $\times$ 0.27	&	-8.9 	&	0.9 	\\
2005.35 	&		0.9 	&	0.79 $\times$ 0.51	&	  4.2 	&	1.2 	\\
2005.54 	&		1.2 	&	0.64 $\times$ 0.28	&	-15.8 	&	1.0 	\\

\enddata
\tablecomments{(1) Observing Epoch, (2) Integration time, (3) Major axis and minor axis of the CLEAN beam, 
(4) Position angle of the CLEAN beam, (5) rms noise level of image.\\
}

\end{deluxetable}


\begin{deluxetable}{clrrrrrr}
\rotate

\tablecolumns{8}
\tablewidth{0pc}
\tablecaption{Positions and flux densities of the knots at 22 GHz.}
\tablehead{
\colhead{Epoch} & \colhead{Comp.}   & \colhead{$F$}    & \colhead{$\Delta{x}$} &
\colhead{$\Delta{y}$}    & \colhead{Major axis}   & \colhead{Axial}    & \colhead{PA}\\
\colhead{} & \colhead{}   & \colhead{[Jy]}    & \colhead{[mas]} &
\colhead{[mas]}    & \colhead{[mas]}   & \colhead{ratio}    & \colhead{[deg]}\\
\colhead{(1)} & \colhead{(2)} & \colhead{(3)} & \colhead{(4)} &
\colhead{(5)} & \colhead{(6)} & \colhead{(7)} & \colhead{(8)} 
}
\startdata

2001.20 			&	F$_{\rm K}$	&	0.049 	$\pm$	0.010 	&	0.00 	$\pm$	0.00 	&	0.00 	$\pm$	0.00 	&	0.17 	$\pm$	0.03 	&	0.64 	$\pm$	0.04 	10&	10 	$\pm$	50 	\\
($\chi^2_\nu$=	1.9 	)	&	E5	&	0.0067 	$\pm$	0.0020 	&	0.27 	$\pm$	0.14 	&	0.23 	$\pm$	0.20 	&	\nodata			&	\nodata			&	\nodata			\\
			&	E4	&	0.0046 	$\pm$	0.0019 	&	0.67 	$\pm$	0.01 	&	0.49 	$\pm$	0.20 	&	\nodata			&	\nodata			&	\nodata			\\
\hline																																																								
2002.45			&	F$_{\rm K}$	&	0.051 	$\pm$	0.010 	&	0.00 	$\pm$	0.00 	&	0.00 	$\pm$	0.00 	&	0.16 	$\pm$	0.15 	&	0.00 	$\pm$	0.00 	&	30 	$\pm$	10 	\\
($\chi^2_\nu$=	1.5 	)	&	E5	&	0.0056 	$\pm$	0.0020 	&	0.26 	$\pm$	0.14 	&	0.24 	$\pm$	0.26 	&	\nodata			&	\nodata			&	\nodata			\\
			&	E4	&	0.0016 	$\pm$	0.0006 	&	0.50 	$\pm$	0.14 	&	0.47 	$\pm$	0.26 	&	\nodata			&	\nodata			&	\nodata			\\
\hline																													
2004.80 			&	F$_{\rm K}$	&	0.068 	$\pm$	0.014 	&	0.00 	$\pm$	0.00 	&	0.00 	$\pm$	0.00 	&	0.34 	$\pm$	0.03 	&	0.37 	$\pm$	0.20 	&	50 	$\pm$	9 	\\
($\chi^2_\nu$=	1.2 	)	&	E4	&	0.0078 	$\pm$	0.0028 	&	0.73 	$\pm$	0.22 	&	0.47 	$\pm$	0.31 	&	\nodata			&	\nodata			&	\nodata			\\
\hline																													
2005.05 			&	F$_{\rm K}$	&	0.059 	$\pm$	0.012 	&	0.00 	$\pm$	0.00 	&	0.00 	$\pm$	0.00 	&	0.16 	$\pm$	0.15 	&	0.61 	$\pm$	0.36 	&	-86 	$\pm$	2 	\\
($\chi^2_\nu$=	1.1 	)	&	E5	&	0.0068 	$\pm$	0.0025 	&	0.24 	$\pm$	0.14 	&	0.28 	$\pm$	0.20 	&	\nodata			&	\nodata			&	\nodata			\\
			&	E4	&	0.0086 	$\pm$	0.0029 	&	0.65 	$\pm$	0.14 	&	0.47 	$\pm$	0.20 	&	\nodata			&	\nodata			&	\nodata			\\
\hline																													
2005.35 			&	F$_{\rm K}$	&	0.074 	$\pm$	0.015 	&	0.00 	$\pm$	0.00 	&	0.00 	$\pm$	0.00 	&	0.35 	$\pm$	0.05 	&	0.21 	$\pm$	0.44 	&	62 	$\pm$	13 	\\
($\chi^2_\nu$=	1.3 	)	&	E4	&	0.0040 	$\pm$	0.0019 	&	0.64 	$\pm$	0.26 	&	0.36 	$\pm$	0.39 	&	\nodata			&	\nodata			&	\nodata			\\
\hline																													
2005.54 			&	F$_{\rm K}$	&	0.057 	$\pm$	0.011 	&	0.00 	$\pm$	0.00 	&	0.00 	$\pm$	0.00 	&	0.14 	$\pm$	0.08 	&	0.78 	$\pm$	0.10 	&	50 	$\pm$	10 	\\
($\chi^2_\nu$=	1.4 	)	&	E5	&	0.0100 	$\pm$	0.0023 	&	0.34 	$\pm$	0.14 	&	0.41 	$\pm$	0.28 	&	\nodata			&	\nodata			&	\nodata			\\
			&	E4	&	0.0059 	$\pm$	0.0018 	&	0.72 	$\pm$	0.14 	&	0.34 	$\pm$	0.28 	&	\nodata			&	\nodata			&	\nodata			\\
\enddata
\tablecomments{(1) Observing Epoch, (2) Component names defined as follows;  the knots E4 and E5,  from downstream of the jet to the core F$_{\rm K}$, 
 (3) Flux density, (4) R.A. offset, (5) Dec. offset, (6) Major axis of Gaussian components, (7) Axial ratio of Gaussian components, (8) Position angle of Gaussian components.}

\end{deluxetable}

\onecolumn

\begin{figure}
  \begin{center}
 \epsscale{0.75}
\plotone{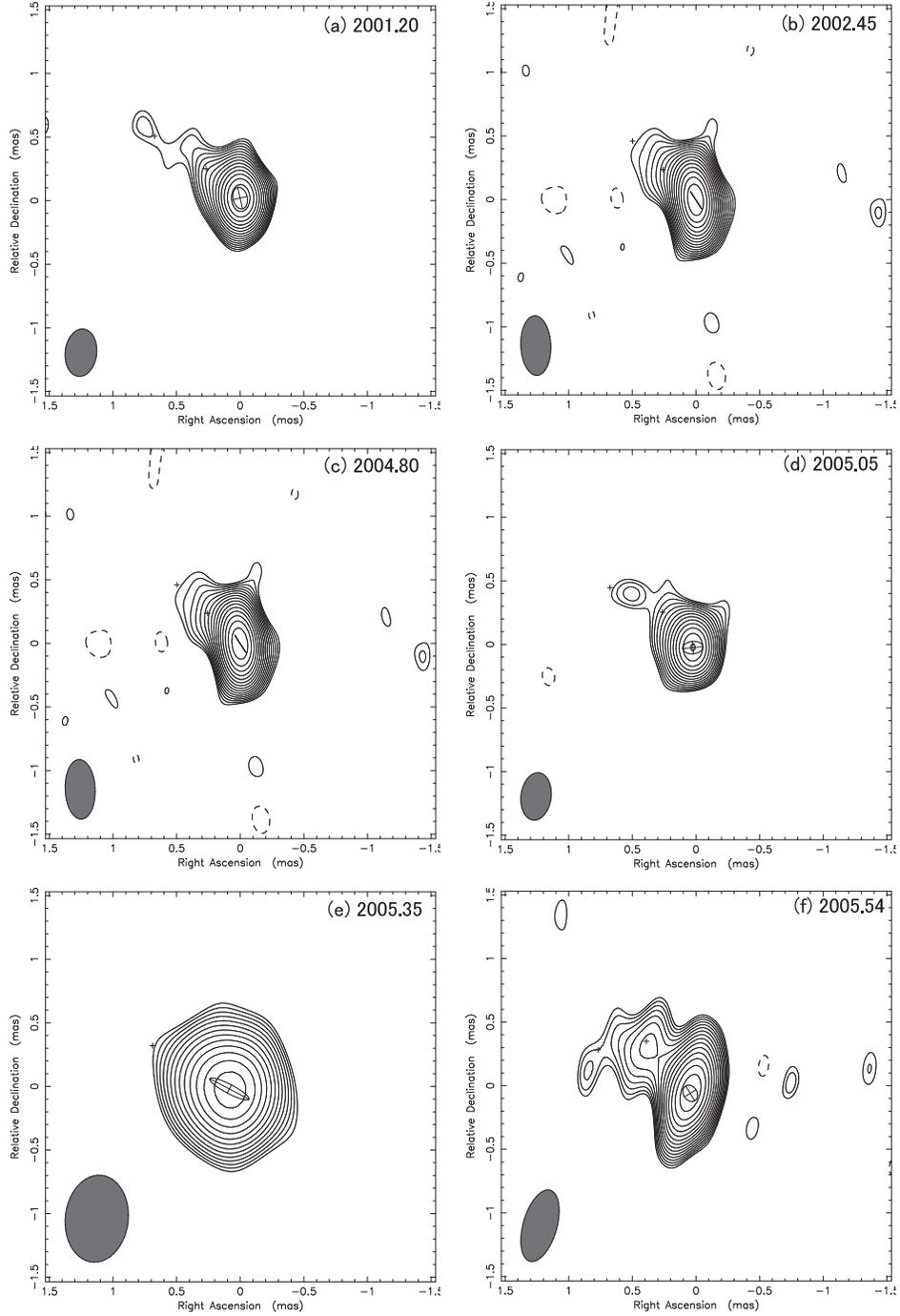}
\caption{Images of 3C 66B. Contours are drawn at $3\sigma \times 1.2^n$ ($n=0, 1, 2, ,,,$). Each value of $\sigma$ is (a) 0.7, (b) 0.9, (c) 0.7, (d) 0.9, (e) 1.2 and (f) 1.0 mJy beam$^{-1}$.
The synthesized beam is shown at the lower-left in each image, and its accurate size is also given in Table 1. The cross marks indicate the positions of knots fitted using a delta function.  
The Gaussian component of the core is also shown.}
 \end{center}
\end{figure}

\begin{figure}
  \begin{center}
 \epsscale{0.75}
 \plotone{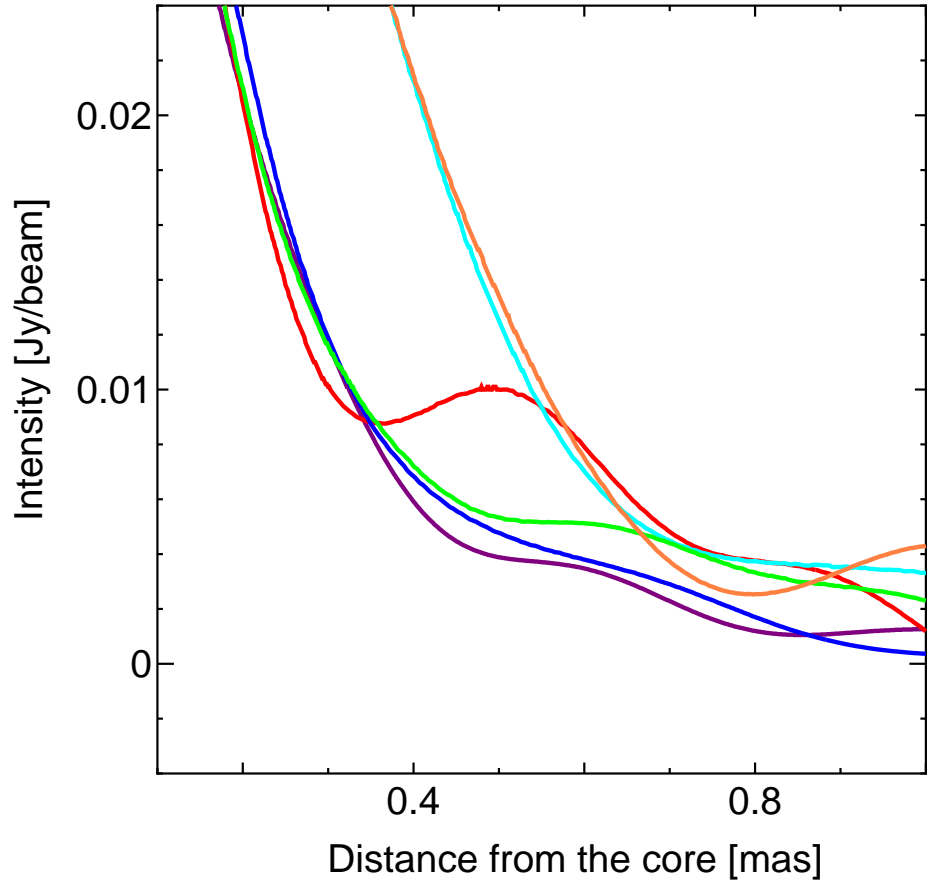} \vspace{4cm}
\caption{Intensity profile along the averaged PA of the jet. The purple, blue, light blue, green, orange, and red lines show the data at epoch 2001.20, 2002.45, 2004.80, 2005.05, 2005.35, and 2005.54, respectively.}
 \end{center}
\end{figure}

\begin{figure}
  \begin{center}
 \epsscale{1.0}
 \plottwo{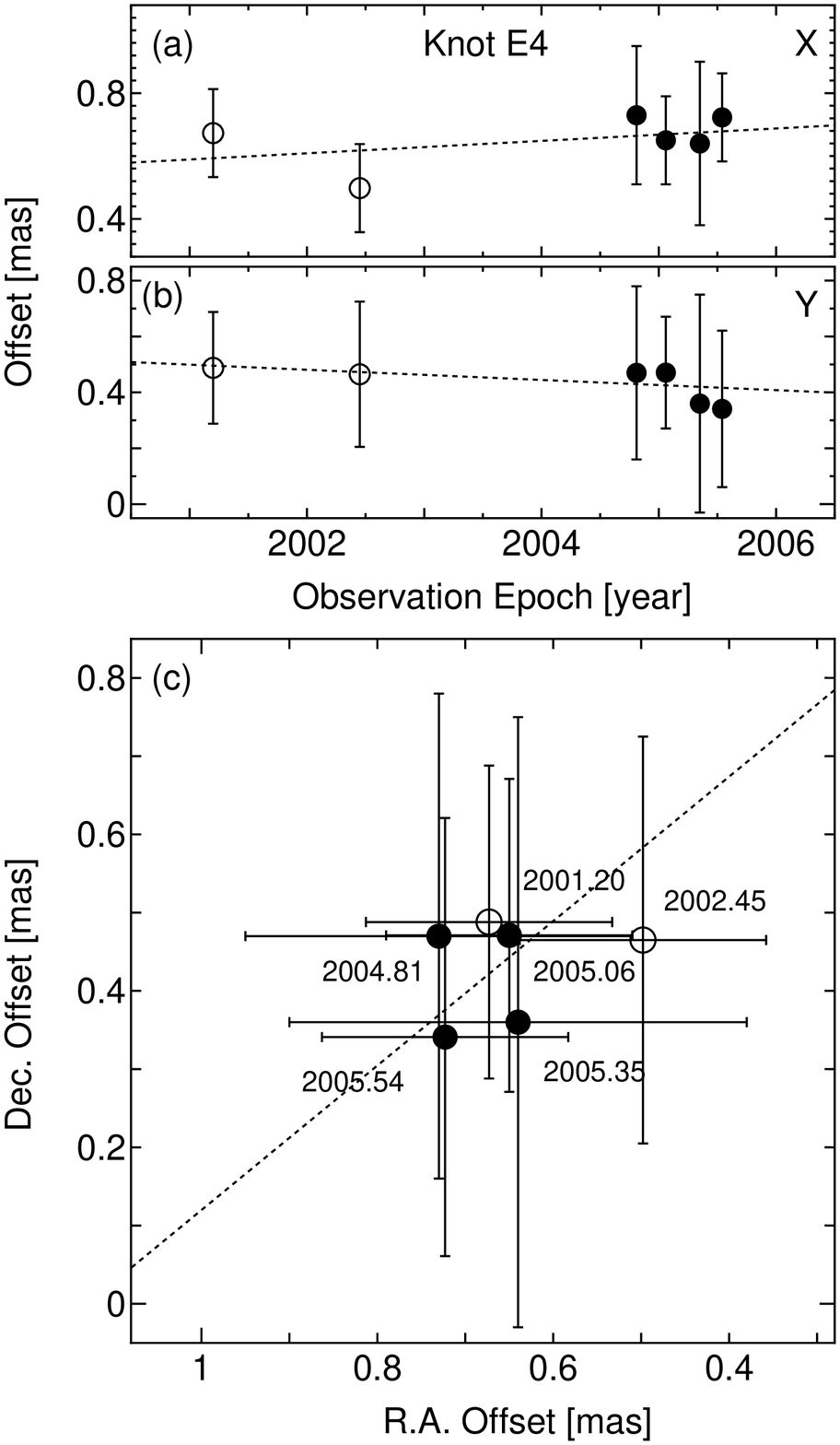}{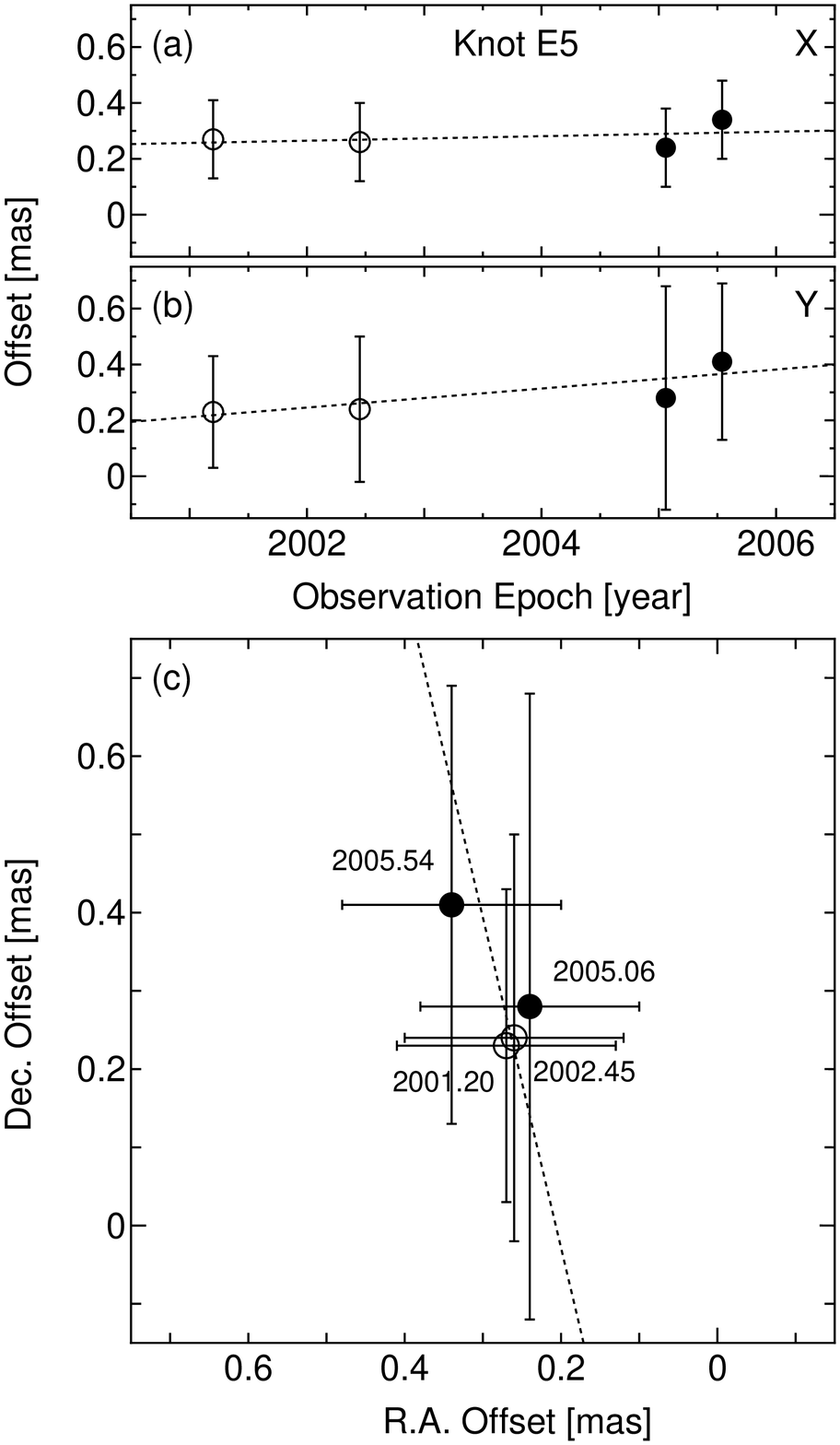}
\caption{Linear fitting of the relative position shift of knots E4 (left) and E5 (right).  (a)  Time evolution in the R.A. direction ($X$), (b) time evolution in the Dec. direction ($Y$), and (c) spatial distribution.
 The open and filled circles show the data for Periods I (epochs between 2001 and 2003) and II (epochs between 2004 and 2006), respectively, according to a previous paper \citep{sudou11}. The dotted lines indicate the best fit.
}
 \end{center}
\end{figure}


\begin{figure}
  \begin{center}
\epsscale{0.5}
 \plotone{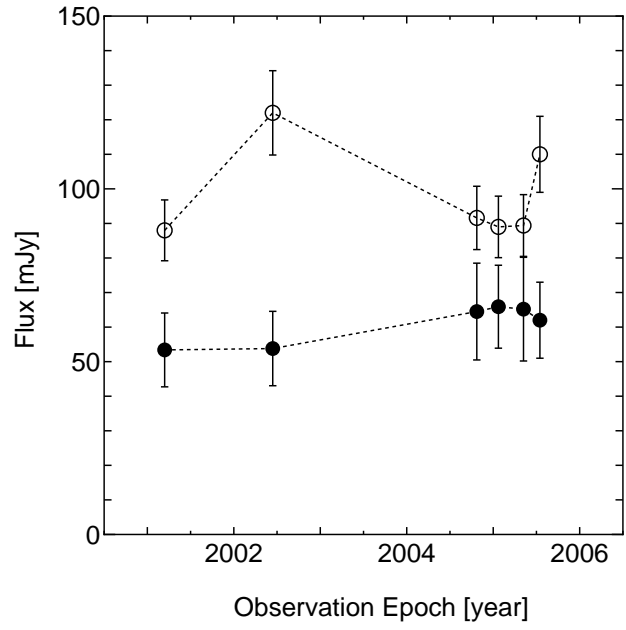}  \vspace{3cm}
  \caption{Comparison of the core flux variation between 8 GHz (open circles) and 22 GHz (filled circles). All images at both frequencies were restored with the 1 mas beam. }
 \end{center}
\end{figure}


\begin{figure}
\begin{minipage}{0.5\hsize}
  \begin{center}
  (a) 8 GHz image\\
  \vspace{5mm}
  \includegraphics[width=6cm,angle=0]{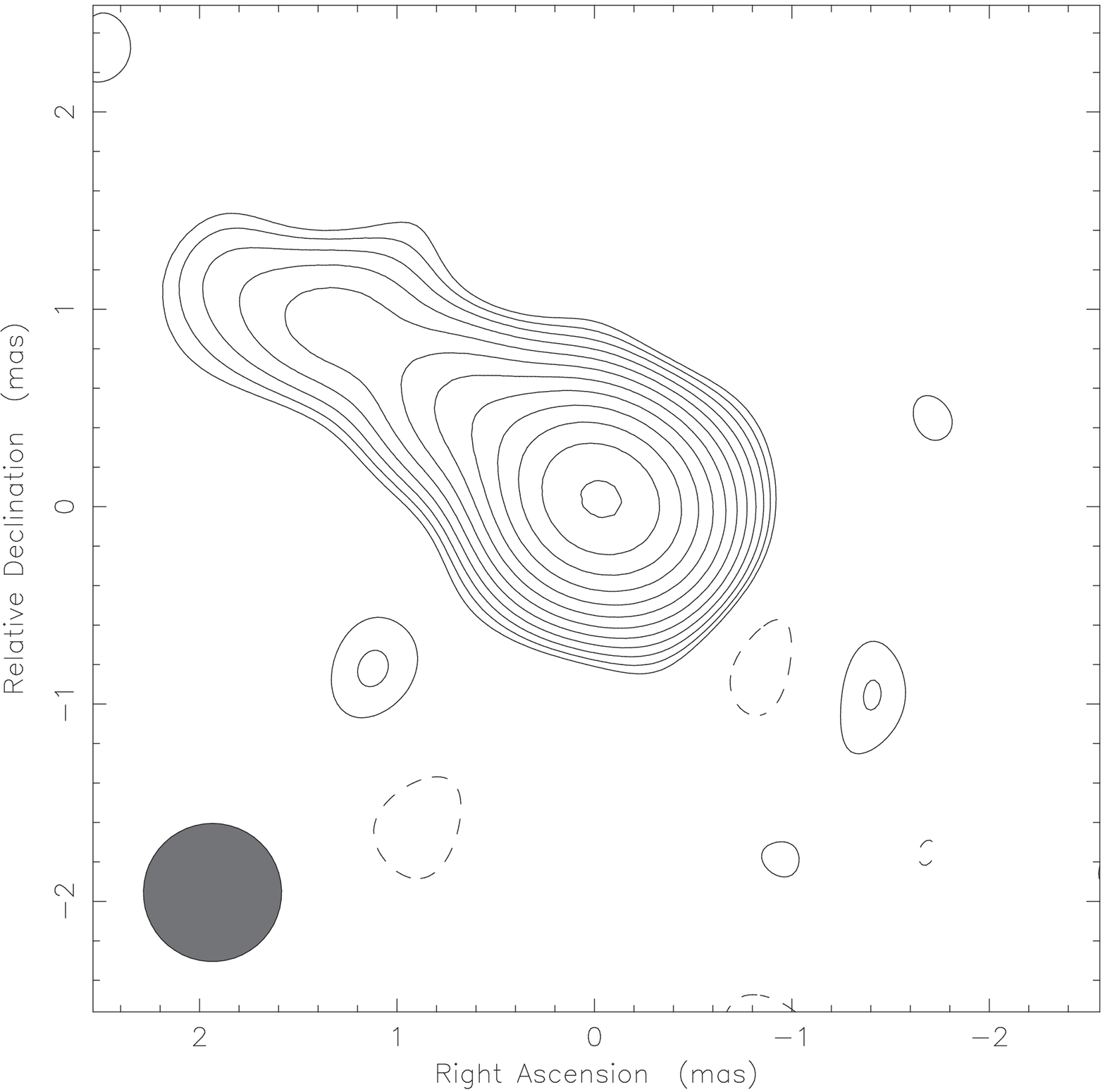}
  \end{center}
  \vspace{10mm}
 \end{minipage}
 \begin{minipage}{0.5\hsize}
  \begin{center}
  (b) 22 GHz image\\
  \vspace{5mm}
  \includegraphics[width=6cm,angle=0]{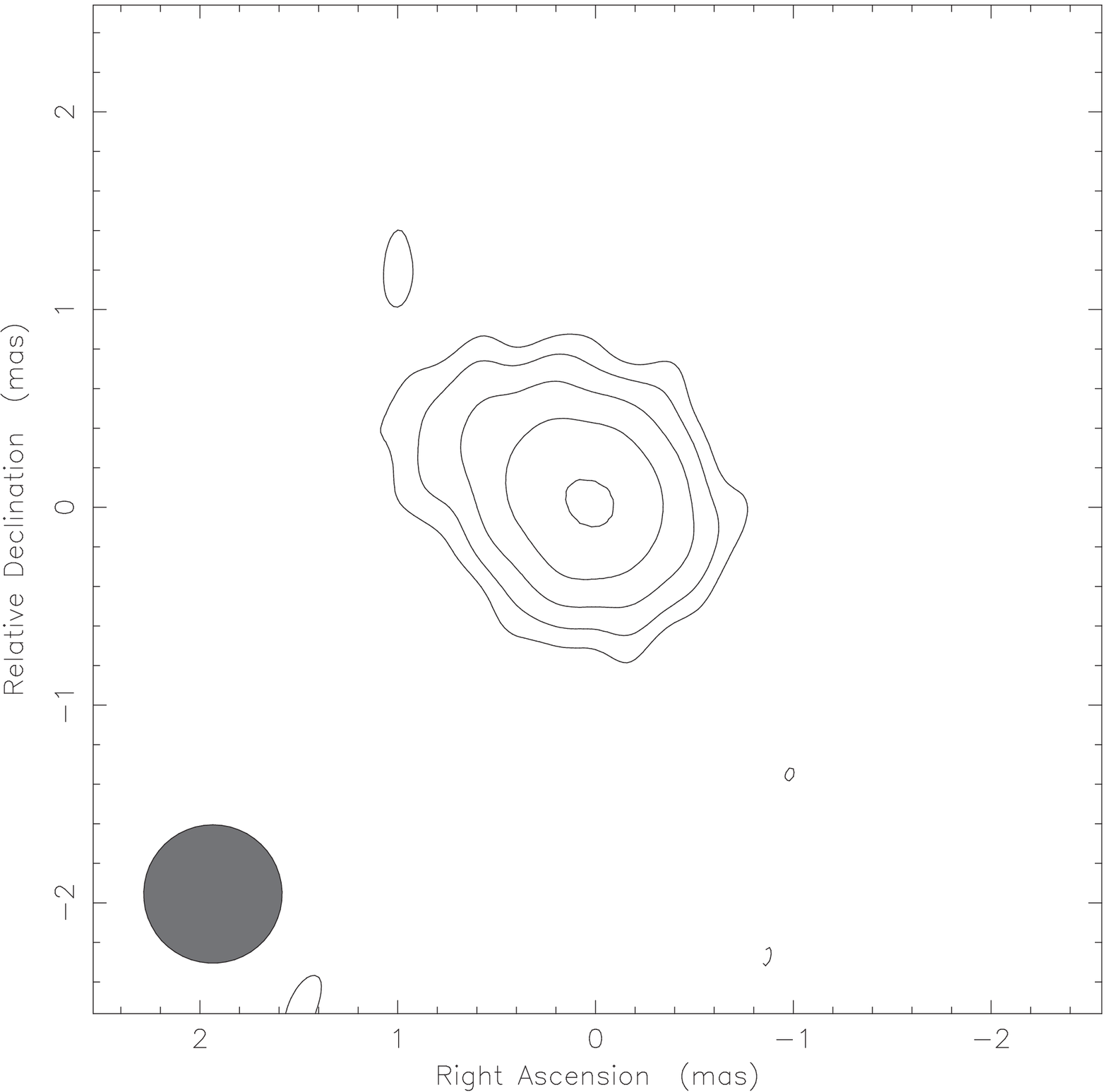}
  \end{center}
  \vspace{10mm}
 \end{minipage}
 \begin{minipage}[t]{0.5\hsize}
  \begin{center}
  (c) spectral index image\\
  \vspace{10mm}
  \includegraphics[width=8cm,angle=0]{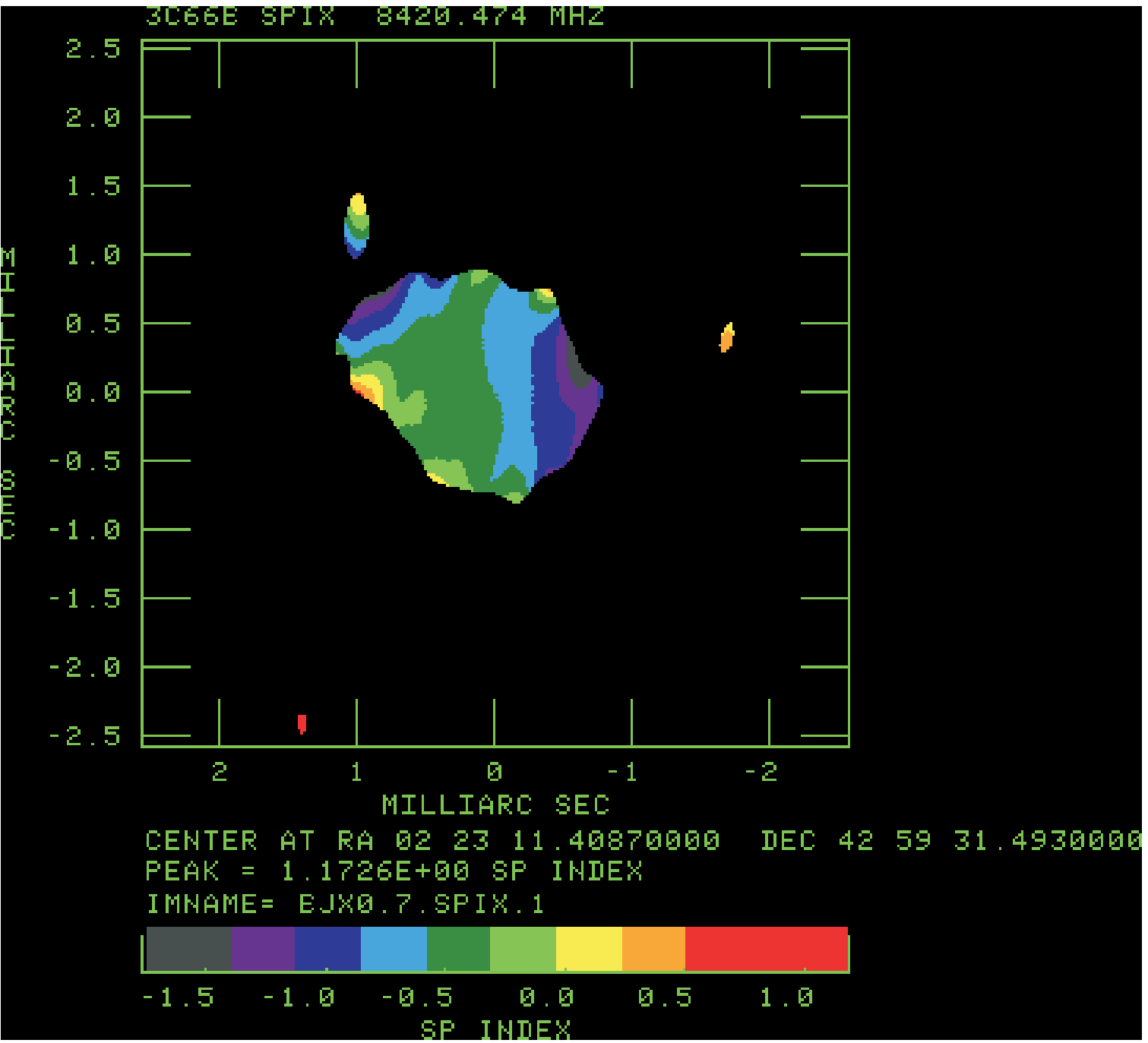}
  \end{center}
 \end{minipage}
 \begin{minipage}[t]{0.5\hsize}
  \begin{center}
  (d) spectral index slice\\
  \includegraphics[width=7cm,angle=0]{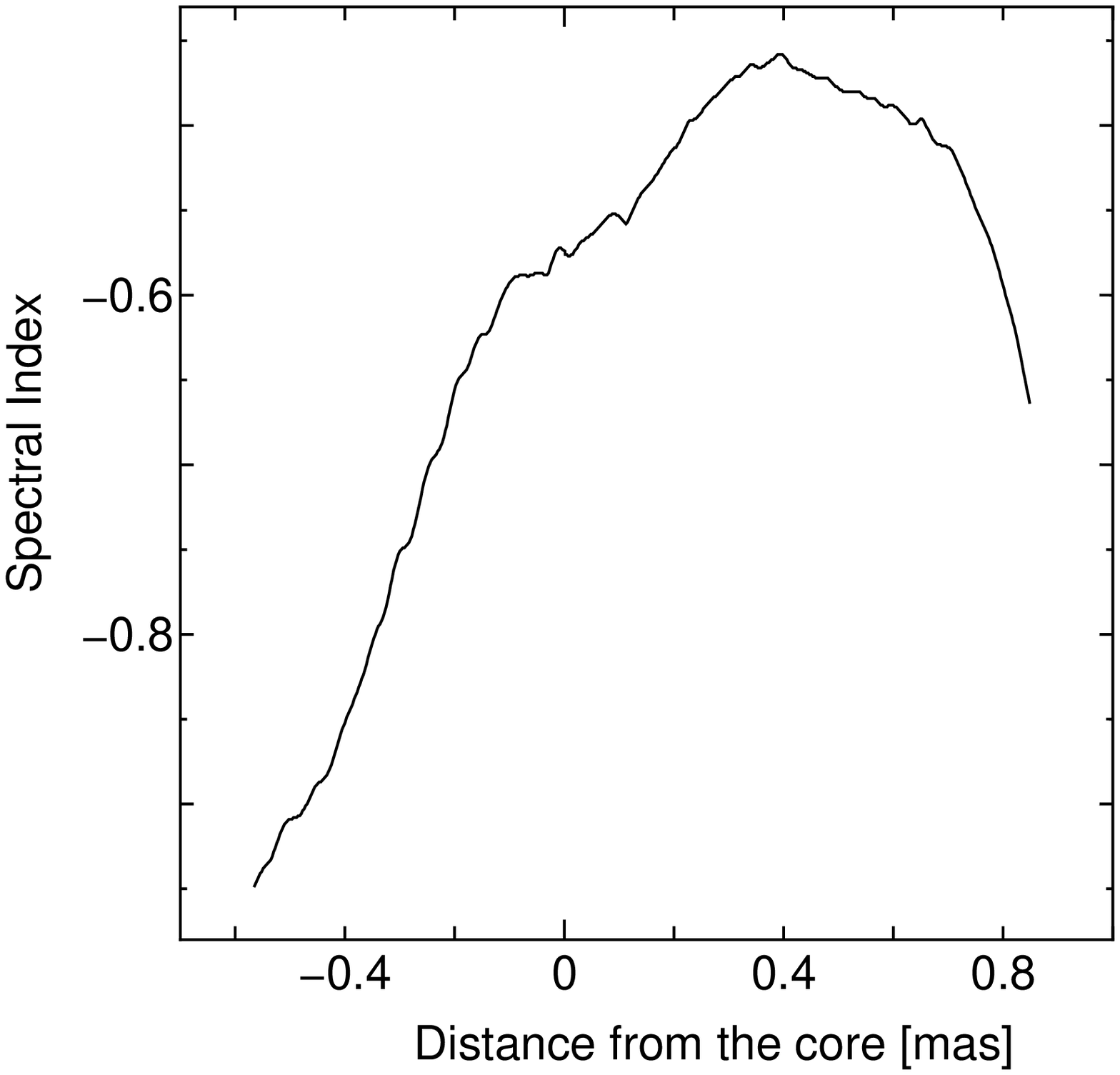}
  \end{center}
 \end{minipage}
   \vspace{10mm}
   \caption{Comparison of the images between 8 and 22 GHz at the epoch 2005.54. (a) 8 GHz image with restoring beam of 0.7 mas, (b) 22 GHz image with restoring beam of 0.7 mas,  (c) spectral index mimage between 8 and 22 GHz, and (d) slice of the spectral index along the jet .}\label{fig3}
\end{figure}

\begin{figure}
 \begin{center}
\epsscale{0.5}
\plotone{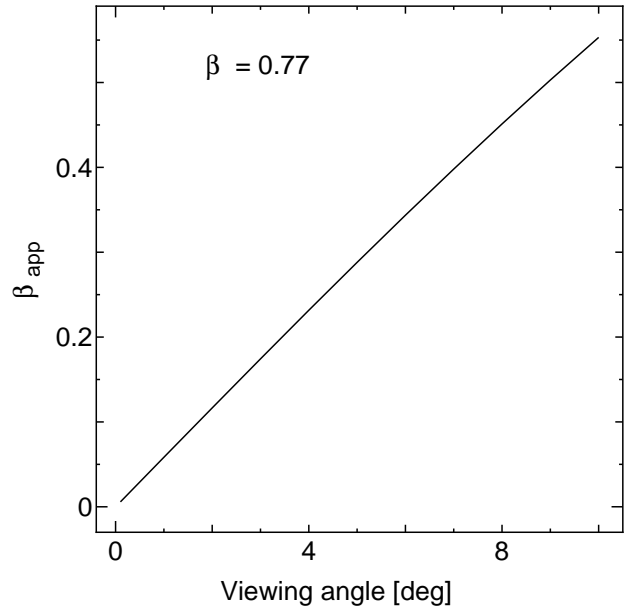}  \vspace{3cm}
  \caption{Relationship between the viewing angle and the apparent velocity of the jet. A jet bulk velocity of 0.77 $c$ is assumed. }
\end{center}
\end{figure}

\end{document}